\documentstyle[11pt,aaspp4]{article}
\input{psfig.sty}

\begin{document}

\title{Detailed Analysis of the Pulsations During and After Bursts from \\
the Bursting~Pulsar (GRO~J1744--28)}

\author{
Peter~M.~Woods\altaffilmark{1,4},
Chryssa~Kouveliotou\altaffilmark{2,4},
Jan~van~Paradijs\altaffilmark{1,3},
Thomas~M.~Koshut\altaffilmark{2,4},
Mark~H.~Finger\altaffilmark{2,4},
Michael~S.~Briggs\altaffilmark{1,4},
Gerald~J.~Fishman\altaffilmark{4}
and W.H.G.~Lewin\altaffilmark{5}
}

\altaffiltext{1}{Dept. of Physics, University of Alabama in Huntsville, 
Huntsville, AL 35899}
\altaffiltext{2}{Universities Space Research Association}
\altaffiltext{3}{Astronomical Institute ``Anton Pannekoek'' and CHEAF,
University of Amsterdam, 403 Kruislaan, 1098 SJ Amsterdam, NL}
\altaffiltext{4}{NASA Marshall Space Flight Center, ES--84, Huntsville, AL
35812}
\altaffiltext{5}{Dept. of Physics and Center for Space Research, 
Massachusetts Institute of Technology, Cambridge, MA 02138}

\begin{abstract}

The hard X-ray bursts observed during both major outbursts of the Bursting
Pulsar (GRO~J1744$-$28) show pulsations near the neutron star spin frequency
with an enhanced amplitude relative to that of the persistent emission. 
Consistent with previous work, we find that the pulsations within bursts lag
behind their expected arrival times based upon the persistent pulsar
ephemeris.  For an ensemble of 1293 bursts recorded with the Burst and
Transient Source Experiment, the average burst pulse time delay
($\Delta$t$_{\rm FWHM}$) is 61.0~$\pm$~0.8 ms in the 25 -- 50 keV energy range
and 72~$\pm$~5 ms in the 50 -- 100 keV band.  The residual time delay
($\Delta$t$_{\rm resid}$) from 10 to 240 s following the start of the burst is
18.1~$\pm$~0.7 ms (25 -- 50 keV).  A significant correlation of the average
burst time delay with burst peak flux is found.  Our results are consistent
with the model of the pulse time lags presented by Miller (1996).

\end{abstract}


\newpage

\section{Introduction}

The Bursting Pulsar, GRO~J1744$-$28, is a low-mass X-ray binary (LMXB) located
on the sky close to the Galactic center (Fishman et al.\ 1995; Paciesas et al.
1996; Kouveliotou et al.\ 1996).  The source exhibits two properties which
separate it from other LMXBs: Type II X-ray bursts (due to spasmodic accretion)
and coherent 0.467 s pulsations (Finger et al.\ 1996).  When GRO~J1744$-$28 was
discovered with the Burst and Transient Source Experiment (BATSE) in December
of 1995 (Fishman et al.\ 1995), it was the only known X-ray burst source to also
emit coherent pulsations (Kouveliotou et al.\ 1996), hence the name the
``Bursting Pulsar.''  Since then, several sources of Type I bursts (due to
thermonuclear flashes) have shown quasi-periodic oscillations, likely connected
to the stellar rotation rate (e.g. Stromayer et al.\ 1996; Strohmayer et al.
1997; van der Klis 1998), and one Type I source, SAX~J1808.4$-$3658, shows
coherent 2.5 ms pulsations in its persistent X-ray flux (Wijnands \& van der
Klis 1998).

During its two years of activity, the Bursting Pulsar produced two distinct
outbursts during which $\sim$ 10,000 hard X-ray bursts were generated.  In
total, more than 10$^{45}$ ergs of energy were released in the form of burst,
persistent and pulsed emission (Woods et al.\ 1999).  The first outburst of
GRO~J1744$-$28 started on 1995 December 2 and lasted until $\sim$ 1996 May 10
(Briggs et al. 1996; Kouveliotou \& van Paradijs 1997) while the second
outburst began on 1996 December 1, and lasted until $\sim$ 1997 April 7 (Woods
et al.\ 1999).  The two outbursts of GRO~J1744$-$28 are similar in many ways. 
After the first day of each outburst, the burst occurrence rate (corrected for
source exposure time) remained constant at roughly 40 events per day.  During
the first 24 hours, the burst rate was much higher at $\sim$ 200 and $\sim$ 135
bursts per day, respectively (Kouveliotou et al.\ 1996; Woods et al.\ 1999). 
For each outburst, the persistent, pulsed and burst flux moved nearly in
lockstep.  The main difference between outbursts was that the persistent,
pulsed and burst flux of the second outburst were all diminished by roughly a
factor of $\sim$ 2 (Woods et al.\ 1999).

Aside from the first day of each outburst, the bursts observed from
GRO~J1744$-$28 did not vary much in duration ($\sim$ 9 s) or in spectral form
(Briggs et al.\ 1996; Giles et al.\ 1996; Kouveliotou \& van Paradijs 1997;
Woods et al.\ 1999).  On the first day, the bursts were typically longer
($\sim$ 15 s).  The burst spectra are consistent with the persistent emission
spectrum, well represented by an Optically Thin Thermal Bremsstrahlung (OTTB)
model with a temperature $kT \sim$ 10 keV.  The constancy of spectra in burst
and persistent emission, the rapid burst recurrence pattern, and other
similarities found with the well known Type II burst source, the Rapid Burster,
suggested that the bursts from the Bursting Pulsar were also Type II events
(Kouveliotou et al.\ 1996; Lewin et al.\ 1996; Kommers et al.\ 1997).  Type II
bursts are due to spasmodic accretion of material onto the surface of a neutron
star caused by some instability within the accretion disk (Lewin, van Paradijs
\& Taam 1995).  In the case of GRO~J1744$-$28, Cannizzo (1996) has proposed a
model where conditions at the inner disk radius of the accretion disk lead to a
Lightman-Eardley instability in the accretion flow, causing the bursts.

Coherent pulsations with a period of 0.467 s were detected from the Bursting
Pulsar as early as 1 December 1995 (Finger 1996), one day before the onset of
burst activity.  GRO~J1744$-$28 has a highly sinusoidal pulse profile in the 20
-- 40 keV energy range with small relative amplitudes of the first two
harmonics: 6.2~$\pm$~0.6~\% and 1.4~$\pm$~0.6~\%, respectively (Finger et al.\
1996). It was realized early into the first outburst that the pulsations
observed from the persistent emission of GRO~J1744$-$28 were also found (at an
enhanced amplitude) during the bursts (Kouveliotou et al.\ 1996).  Using data
acquired with the Oriented Scintillation Spectrometer Experiment (OSSE),
Strickman et al. (1996) showed that the pulses within bursts did not coincide
with their expected arrival times based upon the phase ephemeris of the
persistent emission.  The sense of this pulse arrival discrepancy during bursts
was always in the direction such that the pulses arrived later than expected,
i.e., a pulse time delay.  Furthermore, Strickman et al.\ showed that this
delay reached a maximum ($\Delta$t~$\sim$~90~ms) during the burst interval for
a sample of bursts recorded in December 1995 and January 1996.  This delay did
not recover fully after the burst.  For 10 -- 80 s following the burst,
Strickman et al.\ (1996) found an average residual shift 29~$\pm$~6~ms.  Using
data taken with the Proportional Counter Array (PCA) aboard the {\it Rossi
X-ray Timing Explorer (RXTE)}, Stark et al.\ (1996) found this residual time
lag recovered exponentially on a timescale 700~$\pm$~20~s, although this value
is determined over a lower energy window than OSSE.  The evolution of the peak
and residual pulse time delay through the outburst were investigated using data
taken with BATSE and the PCA. Using BATSE data, Koshut et al.\ (1998) found
that the average pulse time delay for 1.5 sec near the peak of the burst
remained constant both through the first outburst and over the energy range 25
-- 75 keV at $\langle\Delta$t$\rangle_{\rm peak}$~=~74~$\pm$13~ms, despite a
net change in peak flux of $\sim$ 3.3.  Stark et al.\ found that the residual
phase shift (2 -- 60 keV) after bursts changed during the outburst from a time
delay of $\langle\Delta$t$\rangle_{\rm resid}$~$\sim$~20~ms to an advance of
$\langle\Delta$t$\rangle_{\rm resid}$~$\sim$~10~ms.

Using data acquired with BATSE, we have studied properties of the pulsed
emission during and after bursts from both outbursts.  We have tracked the
pulsed amplitude and phase or time delay from the onset of the burst to
$\sim$~240 s following.  We detect modest changes in both the burst and
residual time delay as a function of burst strength.  We find a marginally
significant correlation between the average burst peak flux and the average
time delay within the burst.

\section{Data Analysis}

During the two outbursts of the Bursting Pulsar, BATSE detected 3110 and  2709
bursts, respectively.  For the 7 months between outbursts, both burst  ({\it
RXTE}) and persistent ({\it RXTE} and BATSE) activity were seen intermittently,
but at a much lower flux level (Cui 1998; Stark et al.\ 1998; Kouveliotou \&
van Paradijs 1997).  Of the 3110 and 2709 bursts that were detected with
BATSE, 1350 and 311 triggered the instrument, making available fine time
resolution data sufficient to study the pulsations within bursts.  The reasons
for the decreased number of triggered events for the second outburst were that
the second outburst was dimmer and the trigger criteria were not optimized
until later into this outburst.

For pulse analysis, we used data accumulated with 64 ms time resolution over 4
energy channels covering the range 25 keV to 2 MeV (DISCSC data type).  This
data type provided the largest sample size with an integration time sufficient
to study the pulsations.  Accumulation of this data type begins at 2.048 s
before trigger time and lasts nominally until $\sim$ 570 s beyond trigger time;
however, the data accumulation was decreased during these outbursts to $\sim$
240 s in order to shorten the read-out time period and thereby decrease the
trigger dead-time of the instrument.  Data acquired prior to and after the
DISCSC data readout used for this study, have the same energy binning, but a
coarser time resolution of 1.024 s (DISCLA data type).  Due to the soft
spectrum of the source (relative to classical gamma-ray bursts), source counts
are found only in the first two energy channels, $\sim$ 25 -- 50 keV and 50 --
100 keV.

First, a detector response matrix (DRM) was constructed for each burst.  Based
upon previous spectral analysis of bursts as seen with BATSE (Kouveliotou \&
van Paradijs 1997; Woods et al.\ 1999a), we assumed an OTTB spectral form with
a temperature 10 keV.  Using an arbitrary normalization over two fixed energy
ranges (25 -- 50 keV and 50 -- 100 keV), we folded this model through each DRM
and calculated the expected count rates in the first two channels.  The ratio
of the input energy fluxes to the output count rates provided efficiency
factors that allowed us to convert from channel 1 (channel 2) count rates
directly to 25 -- 50 keV (50 -- 100 keV) energy flux.  Next, a low-order
polynomial (usually less than 3$^{\rm rd}$ order) was fit to approximately 300
s of pre-burst and post-burst data for each channel.  This background model was
subtracted from the time history and the count rates were converted to energy
flux using the efficiency factors.

Pulse timing analysis was not possible for individual events due to an
inadequate signal-to-noise ratio.  In addition, there also exists variablility
on time scales not associated with the stellar rotation, which makes pulse
timing more difficult.  Figure 1 is a selection of bursts (25 -- 50 keV) taken
from the first outburst that show the range of time scale variability involved
within individual bursts.  In order to study the pulse timing within bursts we
were forced to sum bursts together in phase.  This process allows for pulse
timing analysis of the average profile, but at the cost of losing information
about other structures within individual bursts.  The procedure with which we
phase aligned bursts and extracted pulse amplitude and timing information is
described in detail in the next section. 

\begin{figure}[!p]
\centerline{
\psfig{file=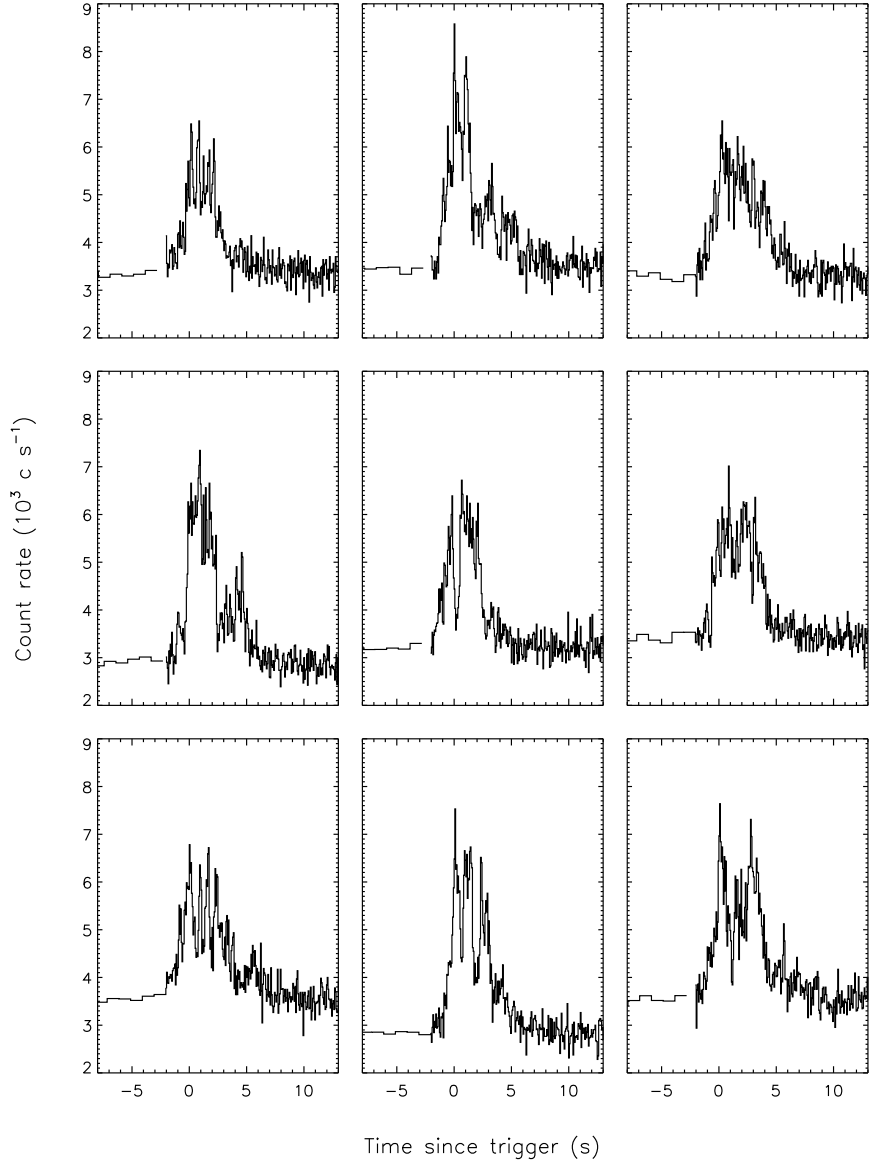,height=7.0in}}
\vspace{0.0in}
\caption{Example burst profiles taken from the first outburst (25 -- 50
keV).  From left to right then top to bottom, these profiles are BATSE triggers
4296, 4325, 4345, 4352, 4359, 4360, 4375, 4382 and 4403.  Time resolution shown
is 0.064 s for time $>$ trigger - 2.048 s, and 1.024 s resolution prior.  
\label{fig:ex_prof}}
\vspace{11pt}
\end{figure}

\subsection{Phase Alignment}

A detailed pulse phase ephemeris was constructed by fitting pulse phases from
non-burst times during the outburst with an orbital model and a quadratic spline
phase model.  The pulse phases were obtained from the fundamental Fourier
amplitudes of pulse profiles derived from BATSE folded-on-board pulsar data in
the 20 -- 40 keV band, which was taken during both outbursts.

Initially, the relative alignment of each burst was determined by the trigger
time and the pulse phase ephemeris.  The first occurrence of zero phase after
the start of the fine time resolution data was found for each event.  Here,
zero phase relates to the sinusoid of the fundamental frequency (2.141 Hz). 
The light curve was shifted to the left by this amount and stored into a
template.  Since the frequency is slightly different for each burst due to
spacecraft orbital Doppler shifts, binary orbital Doppler shifts, spin torques
due to accretion, and the sampling of the data is fairly coarse (0.064 s)
relative to the period of the pulsar (0.467 s), we chose to split each bin when
storing the fluxes in the average burst profile or template.  The flux of each
bin was assumed to be constant across the bin.  Upon applying this procedure
for all events, we were left with a high signal-to-noise quality template.

\begin{figure}[!p]
\centerline{
\psfig{file=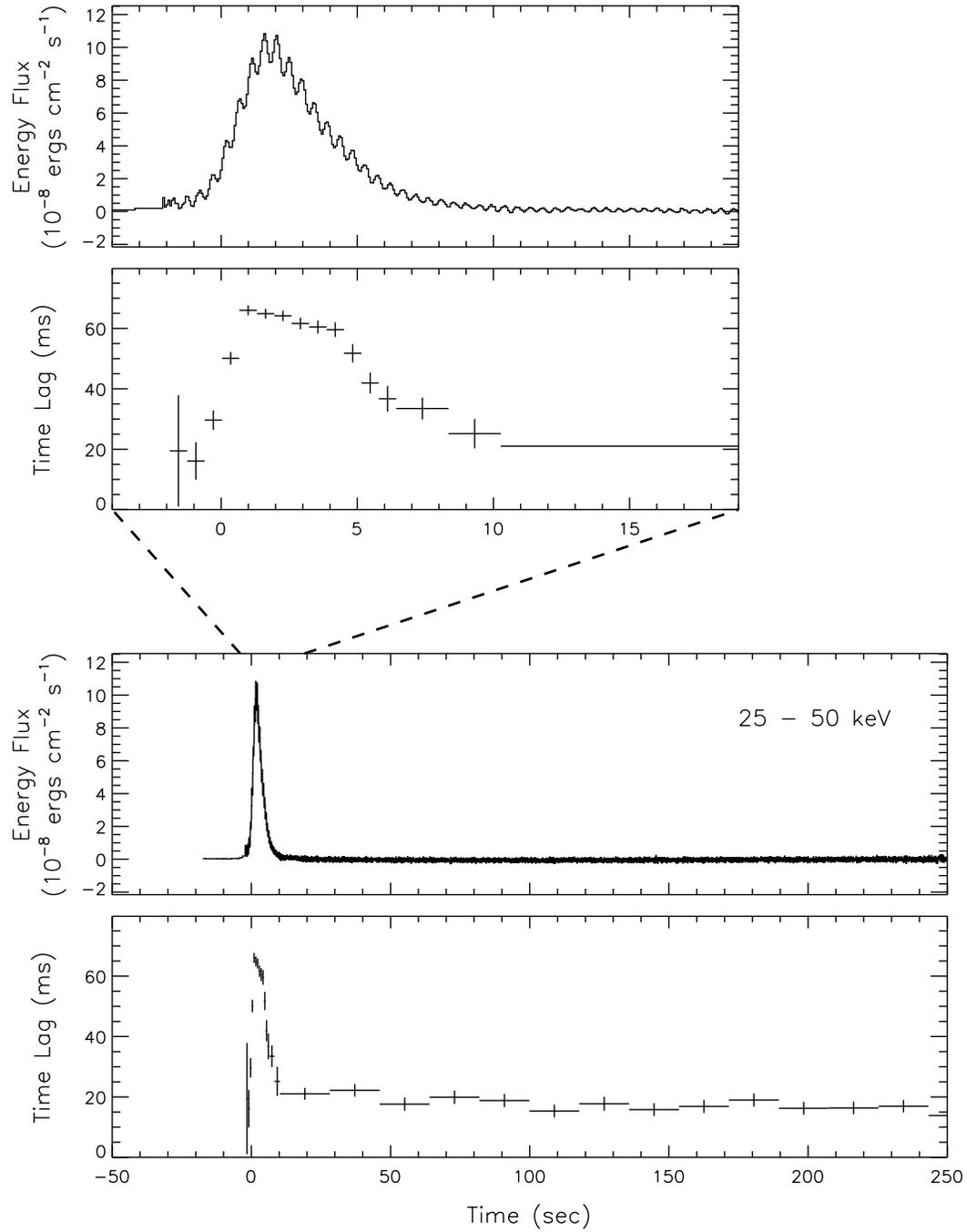,width=6.0in}}
\vspace{-0.2in}
\caption{{\it Top} - Upper panel is the phase aligned burst profile (25 -- 50
keV) of all available bursts with sufficient data coverage from the first and
second outbursts.  Lower panel is the measured time delay versus time
throughout the burst.  {\it Bottom} - Same figure as above, only with an
expanded view to show the residual time delay after the burst.
\label{fig:ch1_prof}}
\vspace{11pt}
\end{figure}

Next, we used this template to refine our alignment procedure.  In order to
optimize the trigger efficiency of the Bursting Pulsar events while retaining
modest sensitivity to GRBs, the trigger criteria were modified during each
outburst.  For the majority of the time, the threshold for a trigger was a
5.5$\sigma$ fluctuation in the second brightest detector integrated over
discriminator channels 1 and 2 (25 -- 100 keV).  For a significant fraction of
the first outburst, this threshold was lowered to 3.5$\sigma$ or 4.0$\sigma$
and the energy range was constrained to 25 -- 50 keV (channel 1).  Even for a
constant trigger criterion, the relative time at which the instrument triggers
within a given burst will change as the burst intensity varies.  As the bursts
become brighter, the instrument triggers earlier into the burst for a fixed
threshold.  Furthermore, changes in the relative position of a burst may occur
for different spacecraft orientations that alter the angle of the source with
respect to the BATSE detectors.  All of the effects mentioned above will
conspire to smear the average profile when aligning each burst relative to
trigger time.  To correct for this, we cross-correlated each event with the
template by shifting the individual burst profile an integer number of cycles
in either direction ($\pm$~6 cycles or $\sim \pm$~3 s) searching for the
optimum alignment.  The cross-correlation was performed only upon the channel 1
rates, due to their superior signal-to-noise. The alignment of the channel 2
data followed from the channel 1 alignment.  Using these refined alignments, we
averaged all `corrected' bursts to create another template and iterated this
procedure several times.

The phase aligned burst profile (25 -- 50 keV) for 1293 bursts from both
outbursts of the Bursting Pulsar is shown in Figure 2.  Figure 3 displays the
phase aligned profile of the same set of bursts, but over a higher energy range
(50 -- 100 keV).  Upon combining data from several events, the various
structures seen in Figure 1 average out leaving a smooth envelope of burst
emission modulated at the neutron star spin frequency.  

\begin{figure}[!htb]
\centerline{
\psfig{file=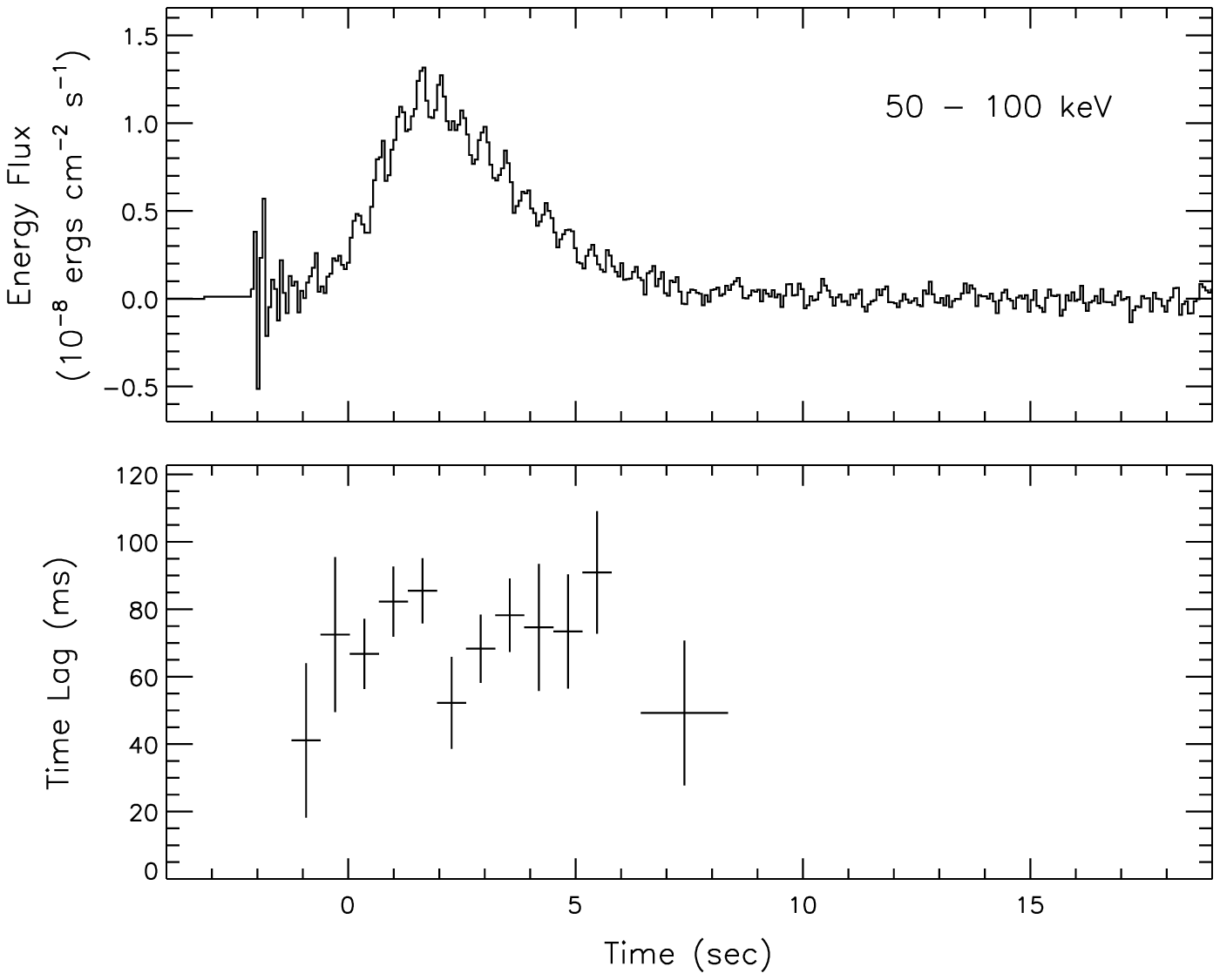,width=5.0in}}
\vspace{-0.2in}
\caption{Phase aligned burst profile (upper panel) over the energy range 50
-- 100 keV, again for the complete ensemble of bursts.  Lower panel displays
the measured time delay through the burst interval.  The statistics were not
sufficient to measure the phase shift outside of the burst interval. 
\label{fig:ch2_prof}}
\vspace{11pt}
\end{figure}

\subsection{Pulse Timing Measurement}

Part of the difficulty in determining the pulse timing or phase shift within
bursts is removing the smooth burst envelope.  The goal is to retain only the
pulsations without inadvertently altering their true phase shift in the
process.  The technique chosen for this purpose was to use a digital high-pass
filter that removes the low frequency power components (upper limit chosen as
1.0 Hz) from the light curve.  One of the artifacts, known as Gibb's
phenomenon, is power leakage at the beginning and end of the filtered light
curve (i.e. loss of the pulsations).  Since the change in time resolution falls
roughly at the beginning of the burst, it is necessary to avoid this artifact
in order to study the pulse timing during the burst interval.  The usual
technique applied during these instances is to pad the background subtracted
light curve with zeroes on either end.  This would be fine for a relatively
flat light curve where padding zeroes would not introduce any large
discontinuities, but at the beginning of the fine time resolution data, the
burst envelope is not yet at the background level.  Simulations using this
filter have shown that a discontinuity in the light curve and/or the slope of
the light curve contributes to power leakage around that point and consequently
alters the phase shift.  To avoid this effect, the pre-burst data (DISCLA) were
saved during the phase alignment procedure and sectioned into phase bins
invoking the same methodology as that described for the DISCSC data
processing.  Of course, no pulsar timing information was available in the
DISCLA data, which have a long integration time (1.024 s); however, these data
were able to provide the necessary padding for removal of the burst envelope.

Upon removal of this burst envelope using the digital filter, only the
pulsations remained.  The next step was to quantify both the pulse amplitude
and phase during and after the burst.  It is clear from the phase aligned
profile that the pulse amplitude varies with time.  We also know from previous
studies (e.g. Strickman et al.\ 1996) and by closer inspection of our data that
the phase shift also varies with time.  Due to the small contributions made by
higher harmonics, the pulse profile is well described by a pure sinusoid. 
Motivated by the observations listed above and our phase alignment procedure,
we chose to fit a sinusoidal function, F({\it t}), with a fixed frequency,
$\nu$ = 2.141 Hz, and a varying pulse amplitude, A({\it t}), and phase shift,
$\phi$({\it t}), given by the following equation.

\begin{displaymath}
   F(t) = A(t) \sin\lbrack2\pi(\nu t - \phi(t))\rbrack 
\end{displaymath}

\noindent  
The choice of time intervals over which fits were performed was determined by
the varying pulse amplitude and signal-to-noise ratio.  During the burst, when
the pulse amplitude was varying rapidly and the signal-to-noise ratio was
greatest, a fit interval of 0.64 s (10 bins) was used.  After the burst was
over and the pulse amplitude was roughly constant, but the signal-to-noise was
reduced, a fit interval of 17.92 s (280 bins) was used.  For the transition
region between, a fit interval of 1.92 s (30 bins) was used.

\begin{figure}[!htb]
\centerline{
\psfig{file=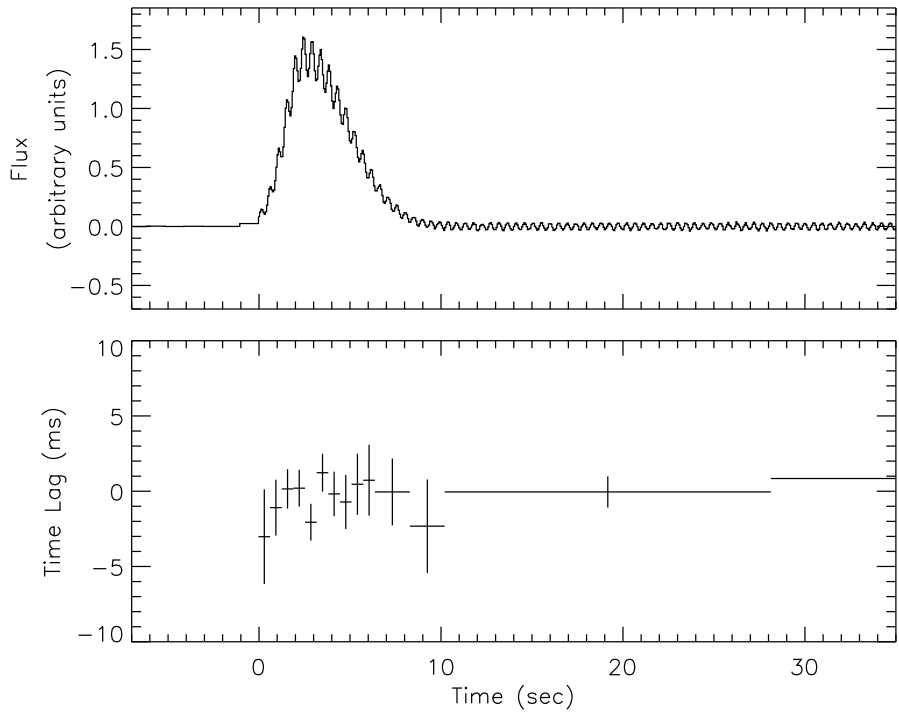,width=5.0in}}
\vspace{-0.2in}
\caption{Simulated burst profile (upper panel) with a constant (zero value)
phase shift model and statistical fluctuations added for the signal-to-noise
found in the 25 -- 50 keV profile of the complete ensemble of bursts (Figure
2). \label{fig:sim_prof}}
\vspace{11pt}
\end{figure}

To ensure that our pulse timing analysis method did not introduce an artificial
phase shift, a number of simulations were performed.  To illustrate the
strength of our method, we describe below an end-to-end test.  We constructed
simulated profiles based upon the observed count rates of the phase aligned
profiles but with known phase shift values.  Figure 4 shows the resulting phase
aligned profile of 1293 simulated bursts with Poisson noise fluctuations
added.  This particular profile has a constant pulse time delay of zero. 
Figure 4 displays the pulse time difference as measured by our method.  As seen
in the bottom panel of Figure 4, no significant time delay is introduced by our
method for this pulse time delay model.  This test was performed for multiple
phase shift models at varying signal-to-noise levels in order to confirm the
accuracy of the pulse time measurement method.

\section{Results}

It is clear from Figure 2 that the pulsed amplitude rises and falls with the
burst envelope.  Figure 5 shows the measured average pulsed flux amplitude as a
function of the average burst flux during each fit interval.  The amplitude is
strongly correlated with the changes in flux during the burst and the pulsed
fraction is systematically larger during the burst rise (diamonds) as compared
to the pulsed fraction during the decline (squares) of the average burst
profile.

\begin{figure}[!htb]
\centerline{
\psfig{file=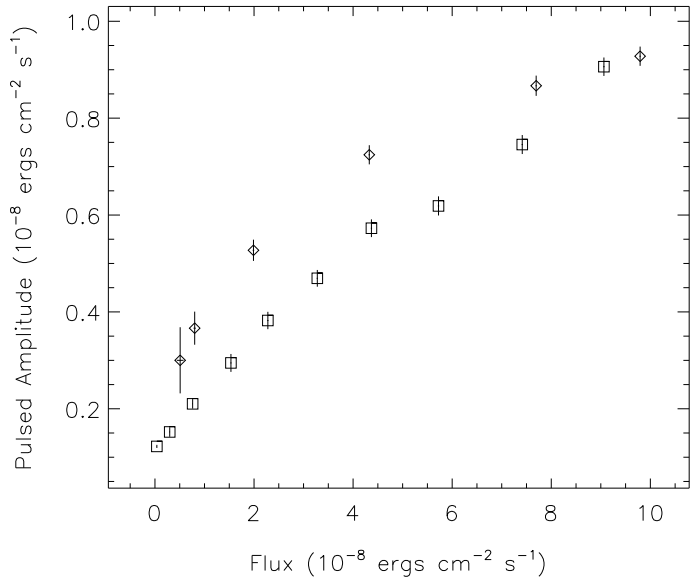,width=3.8in}}
\vspace{0.1in}
\caption{Pulsed flux amplitude versus average burst flux.  Both flux values
given over 25 -- 50 keV.  The diamonds represent measurements during the rise
of the burst and the squares denote intervals after the burst peak.
\label{fig:pf_vs_flux}}
\vspace{11pt}
\end{figure}

As described in the previous section, both the pulse amplitude and the phase
shift were extracted from each fit interval within the phase aligned profiles. 
The 25 -- 50 keV profile shows a quick rise of the phase shift to a maximum
near $\sim$~65 ms and a slower decay to a residual shift of $\sim$~20 ms which
persists to the end of the available data.  A positive shift indicates that the
observed pulse occurred later than expected.  We defined an average burst time
lag ($\Delta$t$_{\rm FWHM}$) to be the average of all time lag measurements
over the full width half maximum (FWHM) interval of the burst profile.  For the
full ensemble of bursts over 25 -- 50 keV, we find $\Delta$t$_{\rm FWHM}$ =
61.0~$\pm$~0.8 ms.  This value is marginally larger for the higher energy band
$\Delta$t$_{\rm FWHM}$ (50 -- 100 keV) = 72~$\pm$~5 ms.  We further defined an
average residual pulse time delay ($\Delta$t$_{\rm resid}$) as the average time
lag between 10 and 240 s.  For the 25 -- 50 keV band, we find $\Delta$t$_{\rm
resid}$ = 18.1~$\pm$~0.7 ms.  The relatively poor statistics in the 50 -- 100
keV band did not allow for a residual time delay measurement.

In addition to averaging over the entire burst sample, phase aligned profiles
were constructed for discrete intervals during each outburst.  The temporal
dynamic range of the available sample of bursts from the second outburst was
insufficient to search for changes in the average burst or residual time delay;
however, the first outburst provided such an opportunity.  Upon analyzing the
data over nine separate time intervals, we found modest, but significant,
changes in both the burst and residual time delay that appear to be correlated
with the rise and fall of the overall outburst.  To better illustrate these
changes, we have grouped the bursts of the first outburst according to peak
flux in the 25 -- 50 keV energy range on the 512 ms time scale.  Nine peak flux
ranges were defined and both the average burst and residual time lags were
measured for each interval (Figure 6; diamonds).  We find a significant
correlation between the peak flux of the average burst profile and the average
burst time delay.  The value of the Spearman rank-order coefficient, $\rho$ =
0.92, corresponds to a 5.1~$\times~10^{-4}$ chance occurrence probability. 
Significant changes are also measured in the residual time delay; however,
there is no significant correlation with peak flux.  When grouped according to
fluence, rather than peak flux, we do not find a significant correlation
between the fluence and the average burst time delay.  For the complete sample
of 1057 events with sufficient data from the first outburst, we find
$\Delta$t$_{\rm FWHM}$ = 61.7~$\pm$~0.7 ms over the energy range 25 -- 50 keV. 

\begin{figure}[!htb]
\centerline{
\psfig{file=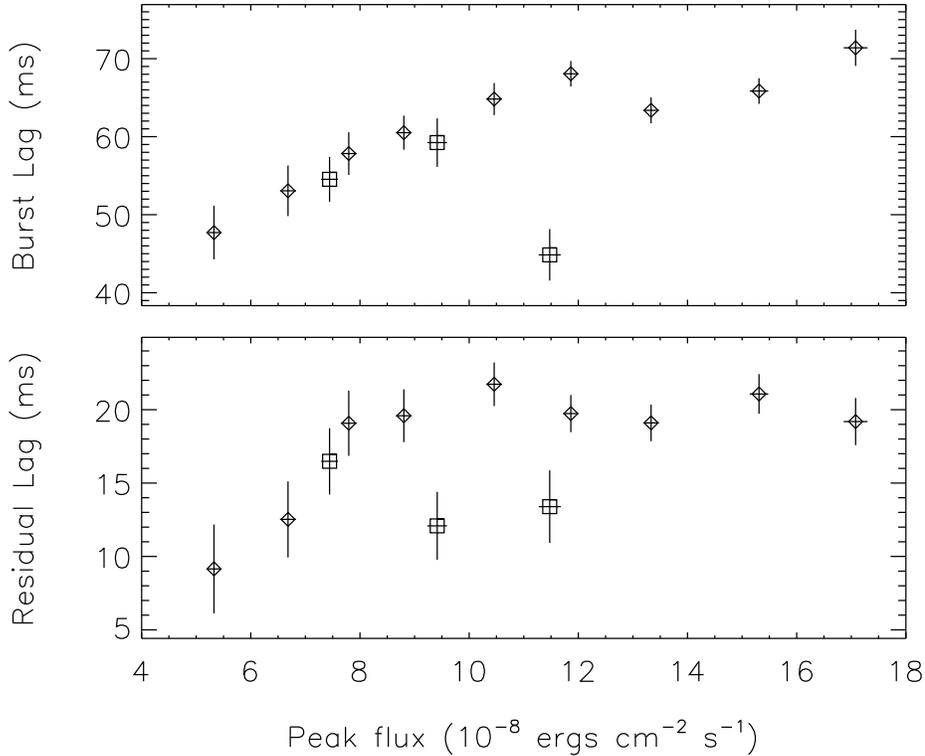,width=5.0in}}
\vspace{0.1in}
\caption{Upper panel -- average burst time delay ($\Delta$t$_{\rm FWHM}$)
versus peak flux of average profile on 64 ms time scale (25 -- 50 keV).  Lower
panel -- residual time delay ($\Delta$t$_{\rm resid}$) from 10 to 240 s past
start of burst versus peak flux of average profile (25 -- 50 keV).  Outburst 1
data is denoted by the diamonds, and outburst 2 by the squares. 
\label{fig:delphi_vs_flux}}
\vspace{11pt}
\end{figure}

Given the limited sample from the second outburst (240 events), we were only
able to split the bursts into three peak flux intervals and still maintain a
reasonable group size.  The average burst time lags of the two lowest peak flux
groups agree well with the values found for bursts of comparable peak flux from
the first outburst (Figure 6; squares).  However, the group with the highest
average peak flux has a significantly smaller average burst time lag as
compared to the first outburst.  For the complete set of bursts from the second
outburst we find $\Delta$t$_{\rm FWHM}$ = 53~$\pm$~2 ms.  The peak flux of the
average profile for the second outburst is only $\sim$ 10\% dimmer than the
peak flux of the first outburst average profile.  The similar values of peak
flux found in these two samples do not reflect an intrinsic similarity in burst
peak flux between the two outbursts, but rather the influence of a variable
trigger selection criterion.  Despite the relatively small difference ($\sim$
10\%) in peak flux of the two aligned profiles from each outburst, they have
significantly different burst time lags due to the contribution from the bursts
of highest peak flux from the second outburst.

\section{Discussion}

The bursts observed from GRO~J1744$-$28 are Type II bursts due to some
accretion instability (Kouveliotou et al.\ 1996; Lewin et al.\ 1996), possibly a
Lightman-Eardley instability (Cannizzo 1996).  Based upon this picture, Miller
(1996) has proposed a model for the pulse time delays.  Due to the 
misalignment of the magnetic and spin axes, only a fraction of the field lines
present avenues for the material from the disk to reach the neutron star (Basko
\& Sunyaev 1976).  This accretion flow geometry leads to a `footprint' in the
form of a narrow arc on the stellar surface which faces the magnetic pole.  For
high accretion rates, the deceleration scale height increases such that the
X-ray emission radiated perpendicular to the accretion flow is enhanced,
creating an `accretion curtain.'  When a burst occurs and the accretion rate
increases, the rapid change in accretion torque may deform the field lines such
that the preferred `pick up' of material from the disk is shifted and
consequently, so is the footprint.  Since the bulk of the emission is directed
perpendicular to the accretion flow, a small azimuthal displacement of this
footprint lends itself to a large observed phase shift of the pulsations. 
Miller suggests the residual phase lag and phase recovery over hundreds of
seconds may be due to a restructuring of the accretion disk.  

The observations presented here have enhanced our knowledge of the pulsed
emission behavior during and after bursts from the Bursting Pulsar.  The strong
correlation between the changes in the unpulsed burst emission component and
the pulsed flux amplitude agrees well with the idea that the height of the
deceleration region scales with the mass accretion rate.  As the scale height
of the deceleration regions grows, so does the emitting area perpendicular to
the accretion flow which leads directly to an amplification of the pulsed
component.

The sign of the phase shift and the correlation of the average burst time lag
with burst peak flux during the first outburst are also consistent with the
accretion curtain model.  As the peak accretion rate increases, one would
expect the magnitude of the azimuthal shift of the footprint or peak amplitude
of the pulse time delay to vary accordingly.  The dependence of the magnitude
of the pulse time delay on burst luminosity was not addressed in Miller's
model.  Given the large azimuthal rotation of the accretion column and the
likelihood of a non-zero angle between the disk plane and the magnetic moment
of the neutron star, one would expect the relationship to be non-linear.  The
observed trend (during the first outburst) is that the burst time lag increases
with burst peak flux up to $\sim$1 $\times$ 10$^{-7}$ ergs cm$^{-2}$ s$^{-1}$. 
Above this burst peak flux value, the average burst time lag flattens out.  It
is not clear why the brightest bursts of the second outburst do not obey the
relationship established during the first outburst.  Incidently, we note that
all of the bursts contained within this subset from the second outburst were
recorded either during or after the 17 day interval where the burst/persistent
OTTB temperature dropped by 20\% (Woods et al.\ 1999).

\acknowledgments{\noindent {\it Acknowledgements} -- PMW acknowledges support
under grants NAG 5-3003 and NAG 5-4419 and the cooperative agreement NCC 8-65. 
CK acknowledges support under grant NAG 5-4799.  JvP acknowledges support under
grants NAG 5-2755 and NAG 5-3674.  MHF acknowledges support under grants NAG
5-4238.  WHGL gratefully acknowledges support from NASA.}

\end{document}